# Thermal Stability of Vapor-Deposited Stable Glasses of an Organic Semiconductor

Diane M. Walters[†], Ranko Richert[‡], M. D. Ediger[†*a]

[†]*Department of Chemistry, University of Wisconsin – Madison, Madison, Wisconsin 53706, USA*

[‡]*Department of Chemistry and Biochemistry, Arizona State University, Tempe, Arizona 85287, USA*

Vapor-deposited organic glasses can show enhanced kinetic stability relative to liquid-cooled glasses. When such stable glasses of model glassformers are annealed above the glass transition temperature $T_g$, they lose their thermal stability and transform into the supercooled liquid via constant velocity propagating fronts. In this work, we show that vapor-deposited glasses of an organic semiconductor, N,N′-Bis(3-methylphenyl)-N,N′-diphenylbenzidine (TPD), also transform via propagating fronts. Using spectroscopic ellipsometry and a new high-throughput annealing protocol, we measure transformation front velocities for TPD glasses prepared with substrate temperatures ($T_{Substrate}$) from 0.63 to 0.96 $T_g$, at many different annealing temperatures. We observe that the front velocity varies by over an order of magnitude with $T_{Substrate}$, while the activation energy remains constant. Using dielectric spectroscopy, we measure the structural relaxation time of supercooled TPD. We find that the mobility of the liquid and the structure of the glass are independent factors in controlling the thermal stability of TPD films. In comparison to model glassformers, the transformation fronts of TPD have similar velocities and a similar dependence on $T_{Substrate}$, suggesting universal behavior. These results may aid in designing active layers in organic electronic devices with improved thermal stability.

## I. INTRODUCTION

Organic glasses are widely studied and are used in a variety of applications. Glasses are advantageous in part because they are non-equilibrium materials and their properties can be tuned depending on the route of preparation.[1–3] In particular, physical vapor deposition can prepare organic glasses with exceptional thermal stability if the deposition conditions are chosen appropriately;[4–6] transformation to the supercooled liquid has been observed to occur at temperatures up to 35 K higher than the glass transition temperature ($T_g$) of the liquid-cooled glass.[7] Stable glasses also have high density[6,8–11] and low enthalpy[5–7,12–15], and can exhibit useful anisotropic structures[16] as evidenced by birefringence[10,11,17], dichroism[17], and X-ray scattering.[14,15,18,19] These solid-state properties are lost on transformation to the equilibrium supercooled liquid, so high thermal stability is important for extending the range of applications for these materials.

[a] Corresponding Author: Contact at ediger@chem.wisc.edu. Phone: 608-262-7273

A recent study has shown that several compounds used in vapor-deposited active layers in organic electronic devices can form glasses of high thermal stability[17], but there are unanswered questions about the mechanism by which these materials thermally degrade. The thermal stability of active layers in organic devices is important as demonstrated by observations that device performance degrades after annealing,[20–22] likely due to pinhole formation and loss of anisotropy. Adachi and coworkers have investigated the transformation mechanism of vapor-deposited active layers, and found that films transformed heterogeneously and were best fit by a model in which mobility is highest near the free surface.[23] Stable glass formation is quite general for vapor-deposited organic molecules and occurs over a wide range of substrate temperatures, so the active layers in many organic devices are likely stable glasses. Understanding the transformation mechanism for the glasses that form active layers could advance device design and improve thermal stability.

Work on model glassformers has established that the thermal stability of thin films of vapor-deposited stable glasses is controlled by a different mechanism than liquid-cooled, ordinary glasses.[24–27] When an ordinary glass is annealed above $T_g$, it transforms to the supercooled liquid by a spatially homogenous process. In contrast, when stable glasses of model glassformers are heated, they transform via constant velocity propagating fronts initiated at a free surface or other interface. For thin stable glass films under about a micron in thickness, the front mechanism dominates and completely controls the thermal stability of the film. Transformation fronts have been directly detected or inferred using a wide variety of experimental techniques. Calorimetry,[25,28–30] dielectric,[26,31] and ellipsometric[27] experiments have observed that the transformation time depends linearly on film thickness for thin films, as is expected for constant velocity propagating fronts. Secondary ion mass spectrometry[24,32,33]and ellipsometry[27] experiments have directly observed the transformation fronts. Transformation fronts are believed to result from kinetic facilitation.[34–36] On annealing, molecules in the interior of a stable glass do not have sufficient mobility to rearrange on a reasonable timescale. However, heightened mobility at a free surface[37] or liquid[37] interface enables adjacent molecules otherwise trapped in the glass to join the liquid. Newly created liquid molecules facilitate motion in their neighbors, causing transformation to propagate from the initial interface into the bulk in the form of a front. Front behavior naturally arises in a kinetic Ising model[34] and in calculations using the random first order transition (RFOT) theory of glasses[36]. These calculations reproduce key experimental features, such as the existence of transformation fronts,[34,36] the competition between front and homogeneous transformation mechanisms under different conditions,[34] and front velocities over a range of annealing temperatures.[36]

In this work, we investigate the thermal stability of vapor-deposited stable glasses of an organic semiconductor, N,N′-Bis(3-methylphenyl)-N,N′-diphenylbenzidine (TPD), that has been widely studied as a hole transport layer.[3,17,20,22,38] We build upon a previously developed high-throughput sample preparation scheme in which deposition occurs onto a substrate with an imposed temperature gradient.[39] We demonstrate a new, high-throughput annealing protocol that uses ellipsometry to efficiently monitor the transformation of nearly fifty different stable glasses of TPD over a wide range of annealing temperatures. We perform dielectric spectroscopy to measure the mobility of supercooled TPD and compare the front velocities to those of other glassformers.

We find that the thermal stability of TPD glasses with high kinetic stability is determined by the velocity of propagating transformation fronts. We observe that the front velocity varies by over an order of magnitude for TPD glasses prepared at substrate temperatures between 0.63 and 0.96 $T_g$, and is imperfectly correlated with the density of glass, similar to results previously reported for indomethacin, a model glassformer. Using our new high-throughput annealing protocol, we calculate the activation energies of the transformation fronts for a wide variety of glasses. We show that the transformation front velocities for TPD glasses prepared at different substrate temperatures have the same activation energy. We find that the mobility of the supercooled liquid and the structure of the glass are independent factors in controlling the thermal stability of TPD films. These results may aid in designing organic electronic devices with improved lifetimes.

## II. EXPERIMENTAL METHODS

Glasses were prepared by physical vapor deposition as has been previously described.[39] TPD (N,N′-Bis(3-methylphenyl)-N,N′-diphenylbenzidine, 99% purity, $T_g$ = 330 K), structure shown in the inset of Figure 2b, was purchased from Sigma-Aldrich and used without modification. Crystalline TPD was loaded into a crucible and placed in a vacuum chamber with a base pressure of $10^{-7}$ torr. The crucible was heated to evaporate TPD and maintain a constant deposition rate of 2.2 ± 0.1 Å/s as monitored by a quartz crystal microbalance next to the substrate. TPD vapor condensed on a temperature controlled silicon substrate 18 cm away from the crucible until the condensed film was about 100 nm thick. The precise thickness and deposition rate were subsequently determined from ellipsometric measurements.

Many glasses with the same chemical composition but different substrate temperatures during deposition ($T_{Substrate}$) were prepared on a single sample using a previously described high-throughput sample preparation scheme.[39] A silicon substrate was suspended between two copper fingers in a vacuum chamber. During vapor deposition, a temperature gradient was imposed on the substrate by heating or cooling each copper finger to a different temperature. A typical temperature gradient spanned 100 K over a 3.2 cm substrate. The substrate temperature during the deposition was determined to be accurate to ±2 K by comparing the birefringence and order parameter of the deposited glasses to previously published work.[17] Three different temperature gradient samples were utilized in this work.

Samples were annealed on the ellipsometer using a home-built temperature controlled translation stage.[10] Ellipsometric measurements were made before, during, and after annealing. Measurements were made on a J.A. Woollam M-2000 spectroscopic ellipsometer at three angles (50°, 60°, 70°) over a 245-1000 nm spectroscopic range. The temperature of the stage was accurate to ±1 K over the temperature range used here based on the comparison to several melting point standards. Nitrogen gas was blown over the ellipsometry stage during measurements to control the environment around the sample. Samples were annealed using either an isothermal or a high-throughput annealing protocol. The annealing protocols are described below.

In the isothermal annealing protocol, samples were annealed at a single temperature for a long period of time. First, ellipsometry measurements were performed at room temperature. Then the sample was ramped at 50 K/min to the annealing temperature and held at that temperature for one hour. Ellipsometric measurements were made every 60 seconds during the annealing. Finally, the sample was returned to room temperature with a cooling rate of 50 K/min; this was maintained through 330 K, the $T_g$ of the material. The sample was measured again at room temperature. This annealing protocol is illustrated in Figure 1a below. Only samples prepared with a single substrate temperature were annealed using the isothermal protocol and one location was measured on each sample.

In the high-throughput annealing protocol, samples were repeatedly annealed for two minutes at steadily increasing annealing temperatures and were measured at room temperature between each annealing step. This temperature protocol is illustrated in Figure 2a below. First, ellipsometry measurements were performed at room temperature. For temperature gradient samples, different locations on the sample were glasses prepared with different substrate temperatures. To measure all these glasses, the sample was "mapped". During mapping, the entire sample was scanned using the translation stage to measure about 90 different glasses prepared at 18 different substrate temperatures; these measurements required about 80 minutes. No measureable aging effects were seen during mapping, as expected because the measurements were done far below $T_g$. After mapping, the sample was brought to the annealing temperature for two minutes before being cooled again to room temperature. The heating and cooling rates were 50 K/min for all but one of the samples analyzed, which was ramped at an uncontrolled, but similar, rate. At room temperature, one glass prepared with each substrate temperature was measured, so about 20 minutes were needed to acquire data. The sample was then annealed again for two minutes at an annealing temperature 2K higher than the previous annealing temperature and then measured again at room temperature. This process was repeated until the sample was fully transformed as confirmed by spectroscopic ellipsometry.

After the above annealing steps, samples were heated to $T_g$ + 15 K at 1 K/min and immediately cooled at the same rate to room temperature. This prepared a 1 K/min liquid-cooled, ordinary glass which is a useful reference state for vapor-deposited organic glasses.[17,39] The entire sample (~ 90 spots) was then mapped again using spectroscopic ellipsometry. The densities of the vapor-deposited glasses were calculated relative to the ordinary glasses on a spot-to-spot basis as previously described.[17,39] Ellipsometric measurements of the as-deposited glass and the 1 K/min liquid-cooled glass were fit with a homogenous model, described below, to determine the film thicknesses. Changes in film thickness are inversely proportional to changes in film density because the film adheres to the substrate and does not flow. The relative densities of the vapor-deposited glasses, combined with the data from reference 17, are shown in Figure 8 below. The $T_g$ of TPD was determined by reheating the 1 K/min liquid-cooled glass at 1 K/min and ellispometrically measuring where the thermal expansion of the material changes from that of the glass to that of the supercooled liquid, as has previously been described for model glassformers.[10,39] The $T_g$ for the model glassfomer indomethacin used for comparison was determined by the same method in Ref 39.

Ellipsometric data were fit with three different models similar to those used previously:[27] a homogenous model, a one front model, and a two front model. All models describe the TPD film on a silicon substrate with 2 nm of native silicon oxide. (1) The homogenous model for TPD is a previously described anisotropic absorptive oscillator model.[17] The TPD film is treated as a single layer and the optical constants and film thickness are fit independently for each measurement. (2) The one front model describes the vapor-deposited film as two layers where the optical constants of each layer are fixed and the layer thicknesses can vary. The bottom layer is the stable glass and the top layer is the ordinary glass (or the supercooled liquid if the measurements were made when annealing above $T_g$.) The optical constants of the stable glass were determined by fitting the homogenous model to the first measurement at the measurement temperature. The optical constants of the ordinary glass (or supercooled liquid) were determined by fitting the homogenous model to the final measurement at the measurement temperature after the sample was fully transformed. The location of the interface between the two layers was determined independently for each measurement. (3) The two front model, illustrated in Figure 1a below, describes the TPD film using three layers:

the middle layer has the optical constants of the stable glass and the top and bottom layers have the optical constants of the ordinary glass (or supercooled liquid depending on the temperature during the measurement.)

We note that the as-deposited TPD glasses are birefringent and dichroic, as established in recent work.[17] The birefringence and dichroism of the as-deposited samples produced for this work are highly consistent and consistent with the previously published results. See Figure S1 in Supplementary Material Document No. ______ for the refractive indexes and extinction coefficients of the as-deposited TPD glasses.[40]

As $T_{Substrate}$ approaches $T_g$ from below, the kinetic stability of vapor-deposited glasses markedly decreases.[17,39,41] Both the one front and two front models failed to fit glasses prepared with substrate temperatures from 0.90 to 0.95 $T_g$ and above 0.96 $T_g$ and these data are not reported; this occurs either because fronts do not exist for these glasses or because there is not sufficient optical contrast between the vapor-deposited glass and the ordinary glass for the fronts to be detected with ellipsometry. Glasses prepared with substrate temperatures ranging from 0.63 to 0.90 $T_g$ were best fit with the two front model, as indicated by monotonically changing front heights and the lowest MSE. Glasses prepared at 0.95 $T_g$ and 0.96 $T_g$ were also best fit using the two front model, as indicated by monotonically changing front heights and the lowest MSE (within a few percent.)

Dielectric measurements were performed as has previously been described for similar small organic molecules.[42] TPD crystals were melted and quenched to form a glass. The glass was heated above $T_g$ and the frequency dependence of the dielectric response of the supercooled liquid was measured over a range of temperatures. A dielectric relaxation time, $\tau_\alpha$, was calculated from the peak in the frequency response of ε”.

# III. RESULTS

## *A. Detection of TPD Transformation Fronts with Ellipsometry*

As shown in Figure 1, the thermal stability of TPD stable glasses annealed above $T_g$ is determined by the velocity of transformation fronts. Figure 1a illustrates the isothermal annealing of a vapor-deposited stable glass of TPD. Between the vertical dashed lines, the sample was annealed for one hour at $T_g$ + 10 K while being measured using spectroscopic ellipsometry. Each ellipsometric measurement was fit independently using the “two front” model described above. The two front model, shown schematically in Figure 1a, describes the vapor-deposited film using three layers with fixed optical constants but variable thickness: a top supercooled liquid layer, a middle stable glass layer, and a bottom supercooled liquid layer.

Fitting the two front model to the experimental data finds the heights of the two interfaces between the layers. The interface heights during the annealing of this sample are plotted in Figure 1b and track the progression of the transformation fronts.

Figure 1b shows the presence of two propagating transformation fronts during the annealing of a vapor-deposited stable glass of TPD at 10 K above its $T_g$. One front originates at the free surface and the other front originates at the substrate. Both fronts progress monotonically with time. Since each ellipsometry measurement is fit independently, this is strong evidence that this film is transforming via a propagating front mechanism.

Figure 1c compares the root mean square error (MSE), a measure for the goodness of the fit to the ellipsometric data, for three models used to fit the film. All the models described the as-deposited glass equally well and have the same MSE, but during annealing the two front model provided the best description of the data. This provides further evidence that the film transforms via surface- and substrate-induced fronts. The other models used to fit the data were a one front model and a homogenous model. The one front model described the organic film with just two layers with fixed optical constants: a supercooled liquid layer over a stable glass layer. The homogenous model describes the organic film as just one layer, but allows the optical constants as well as the thickness of that layer to vary. All samples were best fit using the two front model as indicated by monotonically changing front heights and the lowest MSE (within a few percent.)

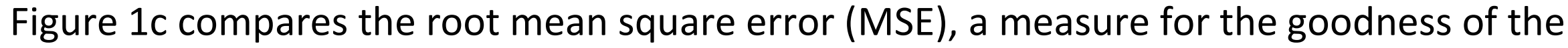

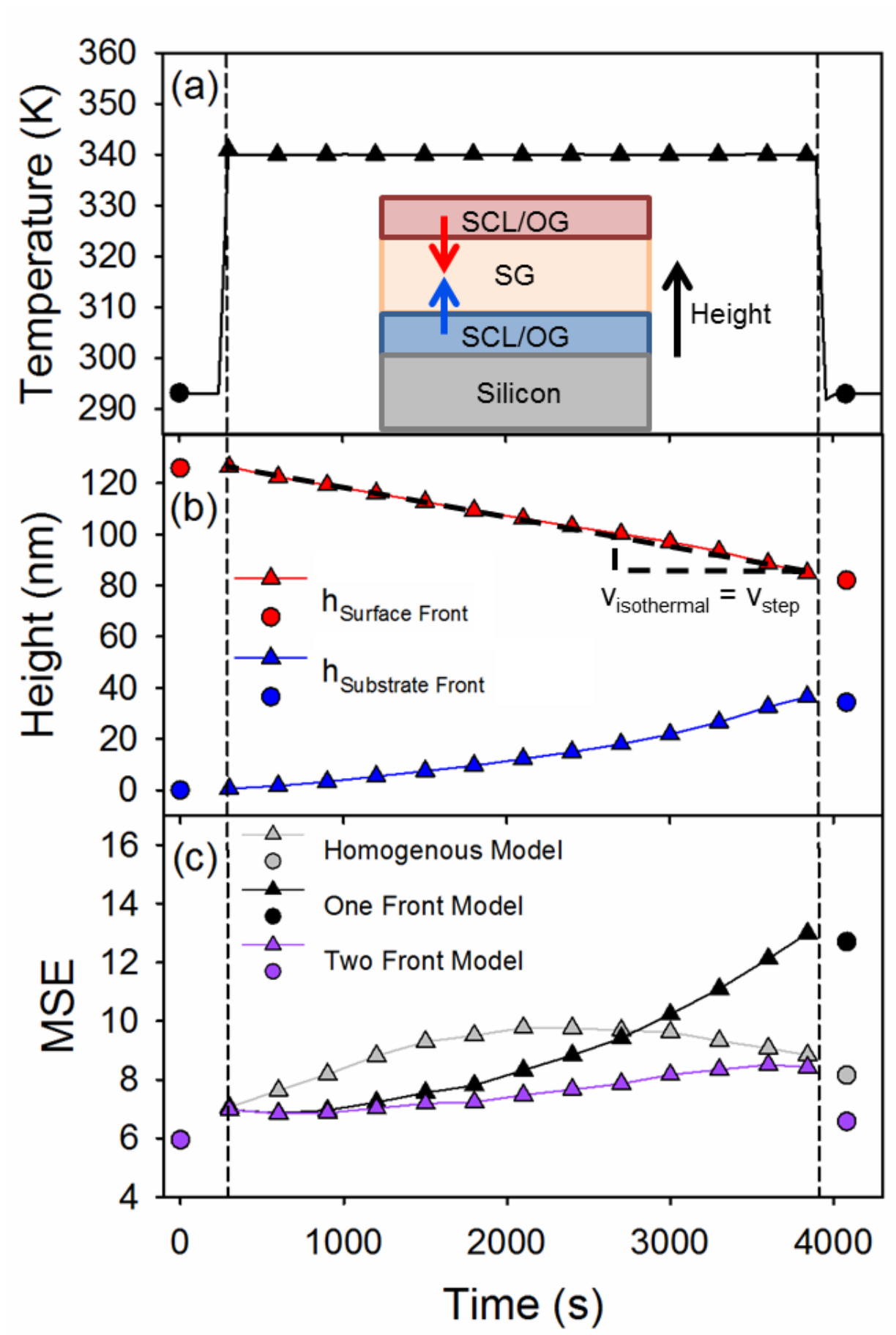

FIG. 1. Isothermal annealing at 340 K of a TPD glass vapor-deposited at $T_{Substrate}$ = 284 K. Vertical dashed lines mark the start and end of the annealing period. Circles denote measurements made at 293 K. Triangles denote measurements made at 340 K. Data density is reduced to 20% for clarity. (a) Temperature profile for the isothermal annealing protocol. Symbols denote ellipsometric measurements. Inset: A schematic illustration of the two front model used to fit ellipsometry measurements. (b) Height of the surface-initiated (red) and substrate-initiated (blue) transformation fronts, as determined by the two front model. (c) Comparison of the root mean square error (MSE) for three different models fit to the ellipsometric data. The two front model has the lowest MSE during the annealing.

The front velocity can be calculated from the ellipsometric measurements shown in Figure 1b. The slope of the triangles indicates the front velocity of the surface-initiated front at the annealing temperature. The same velocity is found by using the difference between the two room temperature measurements (circles) and dividing by the annealing time. Since the front velocities measured by these two methods are equivalent, we can develop a high-throughput annealing protocol in which front velocities are calculated using only measurements made at room temperature before and after annealing, as described in the next section.

### *B. High-Throughput Annealing*

We used a high-throughput protocol to measure the thermal stability of vapor-deposited glasses at many different annealing temperatures in a single experiment, as illustrated in Figure 2. In this protocol, the sample is annealed for many short (two minute) intervals and is measured using ellipsometry at room temperature between each annealing step. Ellipsometric data were fit with three different models and all samples were best fit using a two front model, as described above; data after each annealing step were fit independently. The annealing temperature was increased by 2 K with each annealing step. By annealing the sample for two minutes with 2 K steps in annealing temperature, our step ramp protocol is similar to a continuous 1 K/min ramp.

Figure 2b illustrates the propagation of transformation fronts in a TPD stable glass annealed with our high-throughput protocol. Transformation fronts are initiated at the free surface and substrate and progress monotonically through the film with each annealing step. At low annealing temperatures, there is essentially no change in the heights of the fronts, while at higher annealing temperatures there are larger and larger changes in the front heights during each two minute annealing period. Front velocity is calculated by dividing the front progression during the annealing step by the annealing time, so front velocities at higher annealing temperatures are larger. This will be discussed further in the next section.

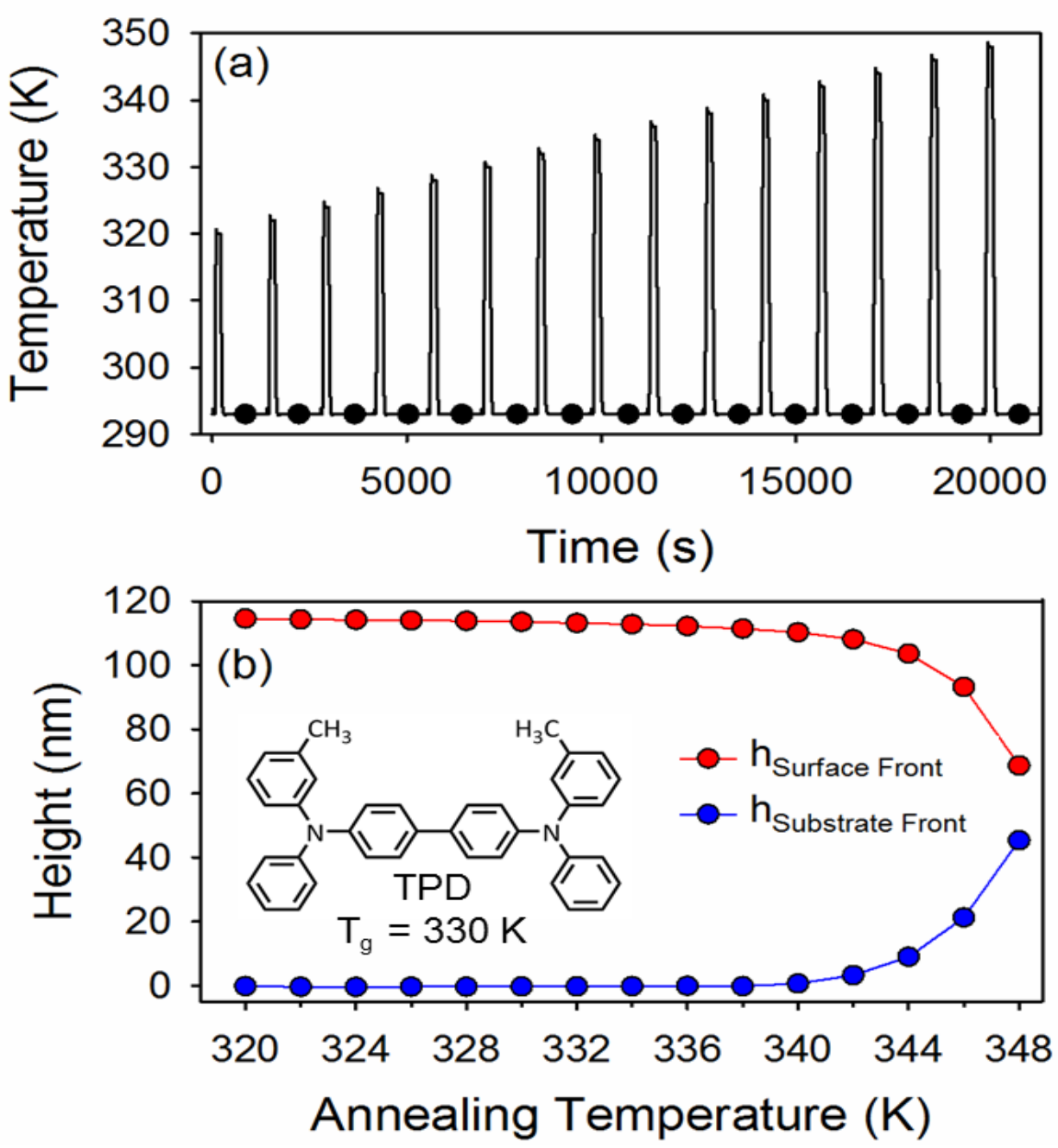

FIG. 2. High-throughput annealing of a TPD glass vapor-deposited at $T_{Substrate}$ = 283 K. (a) Temperature profile of the high-throughput annealing protocol. Ellipsometric measurements are made at 293 K and are denoted with circles. (b) Height of the surface-initiated (red) and substrate-initiated (blue) transformation fronts after each annealing step. Inset: The molecular structure of TPD and its glass transition temperature, $T_g$.

In a single experiment, we measured the thermal stability of many different glasses over a range of annealing temperatures using our high-throughput annealing protocol. Figure 2 illustrates the thermal stability of just one of eighteen different glasses prepared on a temperature gradient sample. Simultaneously measuring the thermal stability of many glasses over a range of annealing temperatures is possible in our protocol because the annealing times are short and the ellipsometry measurements are made at room temperature. By having short annealing times, the sample can be annealed at many different annealing temperatures before it is fully transformed to the supercooled liquid; this allows an efficient survey of the effect of annealing temperature on thermal stability. By making measurements at room temperature after each annealing step, we decouple the measurement time from the annealing time. Long measurement times are needed to measure the many different glasses prepared at different substrate temperatures in order to determine how they changed during each annealing. Room temperature is far enough below $T_g$ for TPD that there are no measureable changes in the properties of the sample over the course of the measurements.

### *C. Influence of Substrate Temperature on Thermal Stability*

As illustrated in Figure 3, vapor-deposited glasses of TPD prepared at different substrate temperatures ($T_{Substrate}$) have different thermal stabilities. Figure 3a shows the propagation of transformation fronts initiated at the free surface for five TPD glasses prepared at different substrate temperatures on the same sample. For each glass, there are greater changes in front height at higher annealing temperatures. Glasses prepared with higher substrate temperatures, up to $T_{Substrate}$ = 283 K, resist transformation most effectively. These glasses have previously been reported to have higher onset temperatures for transformation into the supercooled liquid.[17]

Transformation front velocities for a range of annealing temperatures are plotted in Figure 3b for the five stable glasses of TPD. The front velocities for each glass were calculated by measuring how far the front progressed during an annealing step (shown in Figure 3a) and dividing by the annealing time. Glasses prepared at different substrate temperatures have very different thermal stabilities. For example, when annealed at 340 K, a glass prepared at $T_{Substrate}$ = 209 K (Figure 3b, triangles) has a transformation front that moves nearly 18 time faster than a glass prepared at $T_{Substrate}$ = 283 K (Figure 3b, circles.) For the most kinetically stable glasses near $T_{Substrate}$ = 283 K, the front velocity was only calculated at high annealing temperatures because no measureable front progression was observed at low annealing temperatures. For the least kinetically stable glasses ($T_{Substrate}$ = 209 K), front velocity was only calculated for annealing temperatures just above $T_g$ because the film fully transformed before annealing steps at the highest annealing temperatures were performed. The front velocities for each glass in Figure 3b have very similar temperature dependences, as indicated by the similar slopes. Thus transformation fronts for all of these glasses have the same activation energy, independent of substrate temperature, as will be discussed in more detail in the following section.

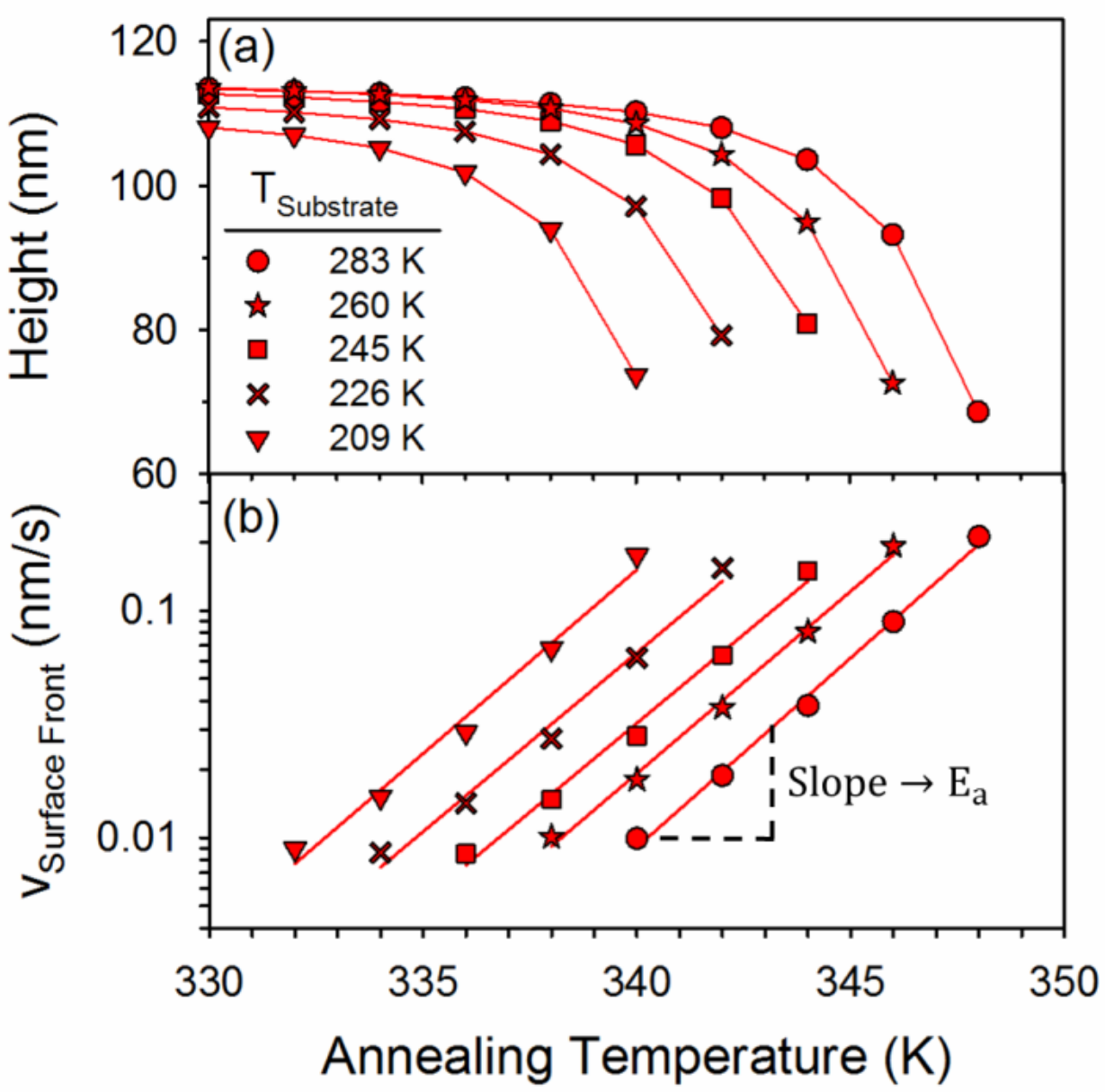

FIG. 3. (a) Height of the surface-initiated transformation fronts after each annealing step in the high-throughput annealing protocol. Different symbols denote glasses prepared at different substrate temperatures on the same sample. (b) Velocities of the surface-initiated transformation fronts in panel (a) during each annealing step. Log ($v_{Surface\ Front}$) increases linearly with the annealing temperature, indicating that the activation energy of the front velocities can be calculated using Equation 1 as described in the text.

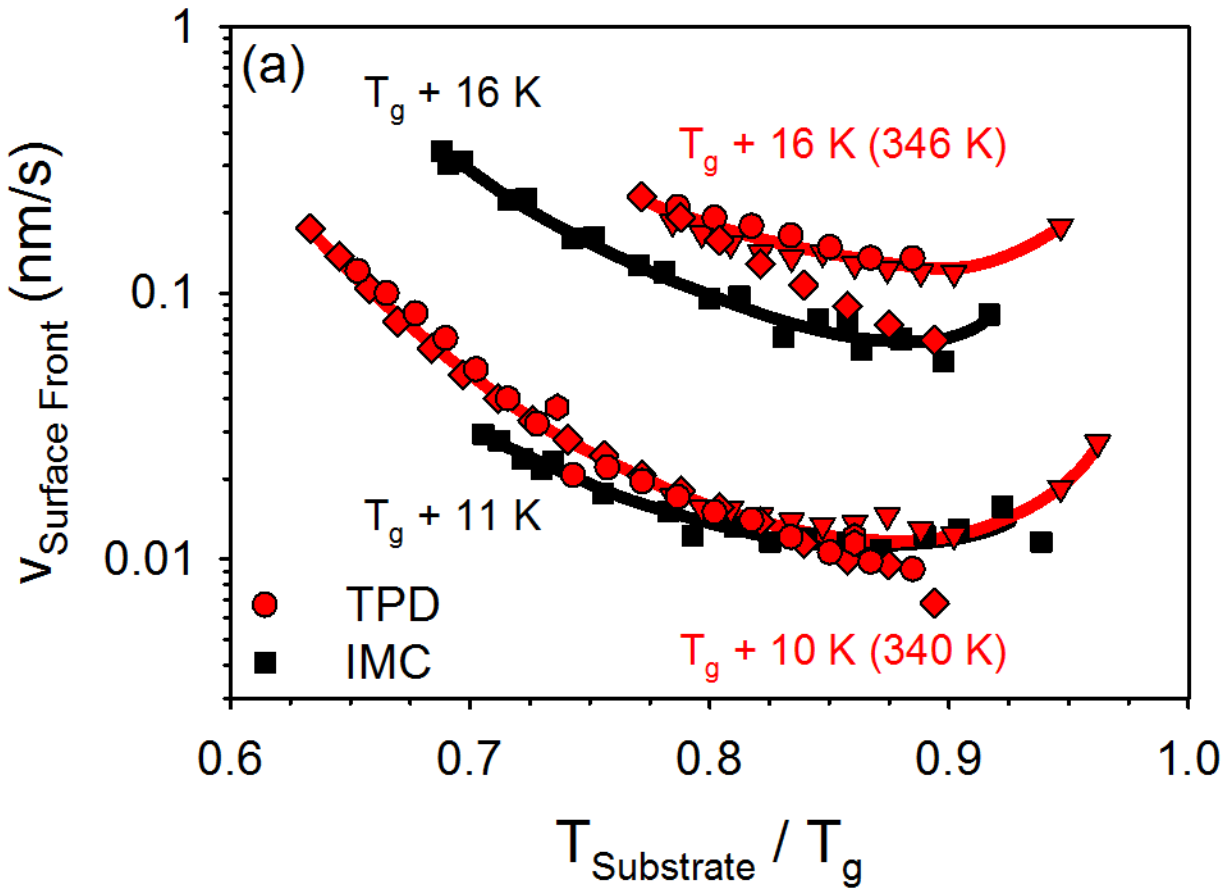

FIG. 4. TPD (red) and IMC (black) surface-initiated transformation front velocities over a range of substrate temperatures for selected annealing temperatures. IMC data are from reference 27. Different symbols represent different samples for TPD. Substrate and annealing temperatures are expressed relative to $T_g$ to compare the two systems.

Figure 4 illustrates how the transformation front velocity for TPD glasses depends upon the substrate temperature during deposition. At two different annealing temperatures, the front velocity varies with substrate temperature by over an order of magnitude. This represents a very significant difference in the thermal stability of these TPD glasses. As shown in Figure 4, front velocities for TPD glasses are similar to those previously measured for indomethacin (IMC, a model glassformer) when comparing the annealing temperatures and the substrate temperatures relative to $T_g$ of each system.[27] In both TPD and IMC, the lowest front velocities (the glasses with highest thermal stability) are produced when the substrate temperature is held near 0.87 $T_g$ during deposition.

Transformation fronts initiated from the substrate have qualitatively the same behavior as fronts initiated from the free surface (substrate-initiated fronts had about 10% larger velocities on average at the annealing temperatures shown in Figure 4), with three important differences. First, as indicated in Figure 1b, the substrate-initiated front does not always propagate at constant velocity. Second, for some glasses on one sample, the substrate-initiated front appeared to show an induction time. Third, different samples showed up to factor of two variation in the velocity of the substrate-initiated front. We do not know to what extent the behavior exhibited at the substrate is an artifact of the fitting procedures. It is possible that modifying the substrate surface prior to deposition would result in more consistent observations. Further discussion will focus on the behavior of surface-initiated fronts.

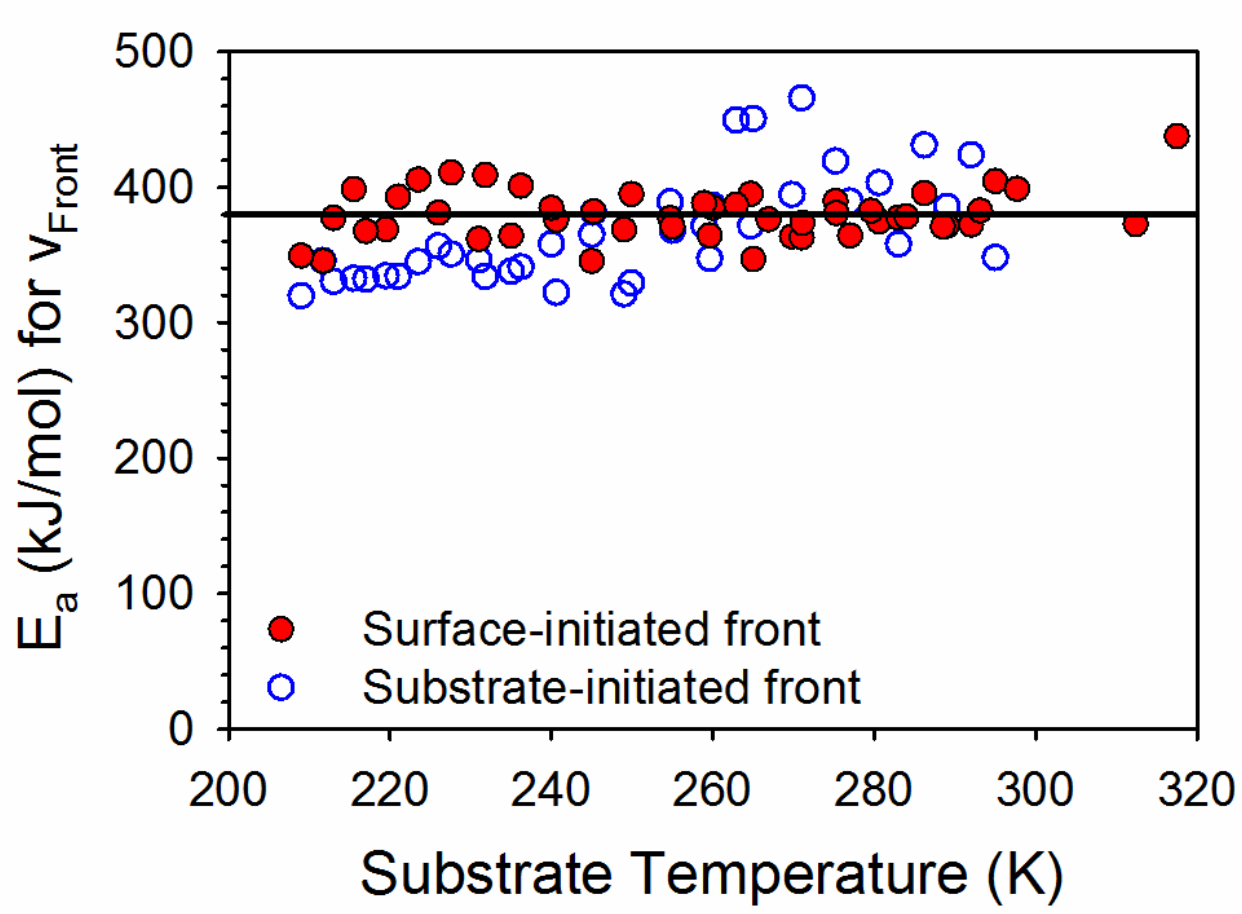

FIG. 5. Activation energies for the surface-initiated (red) and substrate-initiated (blue) transformation front velocities over a wide range of substrate temperatures. A line marks the average activation energy of the surface-initiated fronts, 380 ± 20 kJ/mol.

## D. *Activation Energy of Transformation Fronts*

As shown in Figure 3b, the temperature dependence of the transformation front velocity appears to be independent of substrate temperature. We can test this observation quantitatively by fitting each data set to the Arrhenius equation to extract the activation energy:

$$E_a = -R\frac{\partial \ln v_{Surface\ Front}}{\partial\,(1/T)} \quad (1)$$

Here R is the universal gas constant.

The activation energies for transformation front propagation into vapor-deposited TPD glasses prepared over a wide range of substrate temperature are plotted in Figure 5. The activation energies of the transformation front velocities have no dependence upon the substrate temperature. The surface-initiated front velocity has an average activation energy of 380 ± 20 kJ/mol. Fronts initiated at the substrate had similar activation energies and an average value of 370 ± 30 kJ/mol.

## E. *Dielectric Spectroscopy*

We used dielectric spectroscopy to measure the frequency-dependent dielectric response of the supercooled liquid of TPD over a wide range of temperature. Figure 6a shows the loss component of the dielectric response, $\varepsilon''(\omega)$, which can be represented by Havriliak-Negami fits with $\Delta\varepsilon$ = 0.25, $\alpha_{HN}$ = 0.92, and $\gamma_{HN}$ = 0.29. A characteristic dielectric relaxation time, $\tau_\alpha$, can be calculated from the peak frequency ($f_{max}$) of the loss profile using the following equation:[42]

(2) $\tau_\alpha = 1/(2\pi f_{max})$

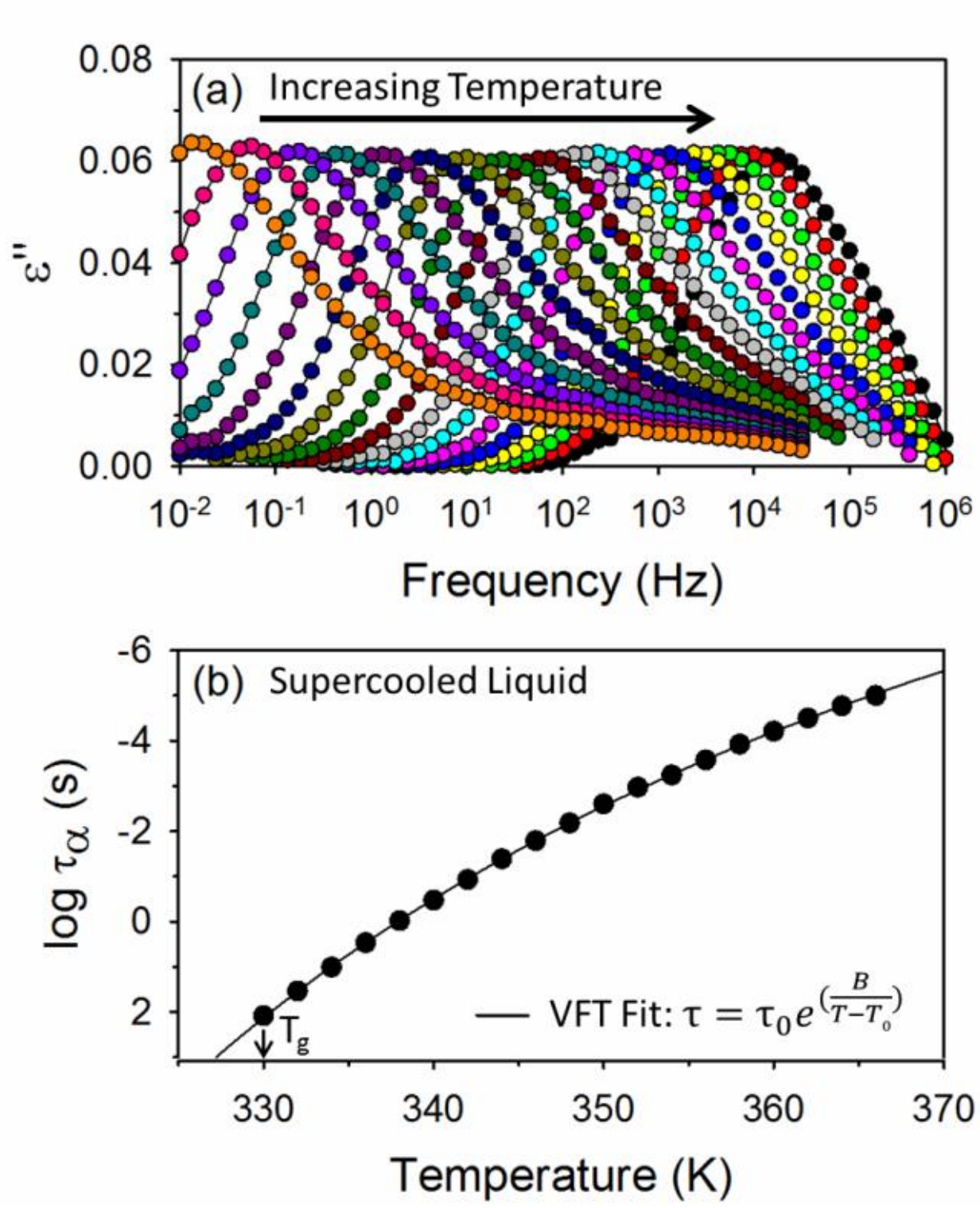

FIG. 6. (a) Dielectric loss spectra for the supercooled liquid of TPD at temperatures spanning 334 K to 366 K with 2 K intervals. The peak in the frequency response shifts to higher frequencies as the temperature increases. (b) Temperature dependence of the dielectric relaxation time $\tau_\alpha$. Data are fit with the Vogel-Fulcher-Tammann (VFT) equation shown.

Figure 6b shows the calculated $\tau_\alpha$ as a function of temperature. The temperature dependence of $\tau_\alpha$ is well-described by the Vogel-Fulcher-Tammann (VFT) equation:[43–45]

(3) $\tau = \tau_0 e^{\left(\frac{B}{T-T_0}\right)}$

The fit parameters for the VFT equation are $\log(\tau_0/s)$ = -19.2, B = 1517.2 K, and $T_0$ = 258.9 K. The $T_g$ calculated from the dielectric relaxation (based upon $\tau_\alpha$ = 100 s) was 330 K. This is in good agreement with the dilatometric $T_g$ of 330 K obtained using ellipsometry and a cooling rate of 1 K/min.[17] The kinetic fragility parameter "m" has a value of 98 for TPD indicating that TPD is a fragile glassformer. We note that the $\tau_\alpha$ values shown for the two lowest temperatures in Figure 6b were determined by an alternate procedure. At these temperatures, the peak of the dielectric loss does not occur in the observed frequency window. The values of $\tau_\alpha$ at these temperatures were obtained by shifting the observed dielectric response to higher temperatures and assuming that the shape of the dielectric response is independent of temperature.

# IV. DISCUSSION

## A. *Universality of Transformation Front Behavior for Glasses with High Kinetic Stability*

Kinetic facilitation,[46–48] the idea that areas with high mobility can induce mobility in neighboring regions of low mobility, offers an explanation for why stable glasses transform into the supercooled liquid via a front mechanism. From this perspective, heightened mobility at a free surface[37] enhances motion in neighboring molecules in a glass. As molecules that border high mobility regions rearrange and equilibrate, they become highly mobile and facilitate further motion. This results in a mobility front that propagates through a glass at constant velocity. Within the framework of kinetic facilitation, propagating transformation fronts are expected to be important for every glass with high kinetic stability that has a high mobility surface or interface.

Our results show that vapor-deposited stable glasses of TPD transform to the supercooled liquid via propagating fronts when annealed above $T_g$. This is illustrated in Figures 1, 2, and 3 for a wide variety of TPD glasses annealed at many different temperatures. All TPD glasses deposited with substrate temperatures ($T_{Substrate}$) between 0.63 and 0.96 $T_g$ transformed by fronts initiated at the free surface and the substrate. Stable glasses can also transform via a bulk mechanism, but the front mechanism dominates in thin films. The film thickness where the front mechanism no longer describes the transformation has been measured for several systems and is on the order of microns.[25,26] The transformation front velocities are the relevant parameter for evaluating the stability of thin TPD films, such as the ~120 nm films in this work.

Transformation via a propagating front appears to be a universal feature in thin glassy films with high kinetic stability, as expected from kinetic facilitation. Figure 7 compares the transformation front velocity of TPD to two model glassformers, indomethacin and ααβ-trisnaphthylbenzene.[32] To fairly compare the different systems, all the glasses shown were vapor-deposited at 0.85 $T_g$. The front velocities in Figure 7 are plotted against the structural relaxation time $\tau_\alpha$ for the supercooled liquid at the annealing temperature in order to account for the different $T_g$ values of the different systems. Figure 7 shows that TPD exhibits transformation fronts similar to model stable glassformers. Among these systems, TPD forms the most stable glasses, as it has the lowest front velocity for a given mobility of the liquid. Figure 4 illustrates that stable glasses of TPD and indomethacin prepared with a wide range of substrate temperatures have similar front velocities when substrate and annealing temperatures are expressed relative to $T_g$.

The high-throughput annealing protocol used here is roughly equivalent to experiments in which temperature is increased at a constant rate and complements temperature-scanning calorimetry measurements by allowing for the direct detection of transformation fronts. Rodríguez-Tinoco et al.[30] and Bhattacharya et al.[29] have investigated stable glasses using calorimetric methods where temperature is rapidly scanned and the excess heat capacity of a stable glass is measured over a broad range of temperatures in a single experiment. Both of these groups observe that the transformation rate is independent of film thickness, which is consistent with transformation via a propagating front mechanism. Rodríguez-Tinoco et al. studied indomethacin and supplemented the fast-scanning measurements with differential scanning calorimetry.[30] They calculated front velocities in quantitative agreement with previously published isothermal annealing experiments and extended the range of investigated annealing temperatures up to $T_g$ + 75 K. Our high-throughput annealing protocol also allows us to access a range of annealing temperatures in a single experiment. Although this range is smaller than in the calorimetry measurements, we access lower annealing temperatures which are likely more relevant for evaluating the thermal stability of molecules used in organic electronics. In addition, ellipsometry experiments directly detect transformation fronts (rather than infer them) and can directly determine the behavior of multiple fronts if they are present.

### *B. What controls the transformation front velocity?*

*Influence of Annealing Temperature.* Within the framework of kinetic facilitation, a propagating transformation front is expected to move more rapidly at higher annealing temperatures because of the higher mobility of the supercooled liquid. Figure 7 is consistent with this idea and further shows that front velocities for the three systems show a similar dependence upon the structural relaxation time $\tau_\alpha$ of the liquid. Using temperature-ramping calorimetry experiments, Rodríguez-Tinoco et al.[30] have shown that the power law relationship between velocity and $\tau_\alpha$ continues for indomethacin over a large temperature range up to $T_g$ + 75 K. It is noteworthy that the front velocities have weaker temperature dependences than $\tau_\alpha$ for all three systems. Front velocities have temperature dependences more similar to that of the supercooled liquid diffusion coefficients for indomethacin and ααβ-trisnaphthylbenzene.[32]

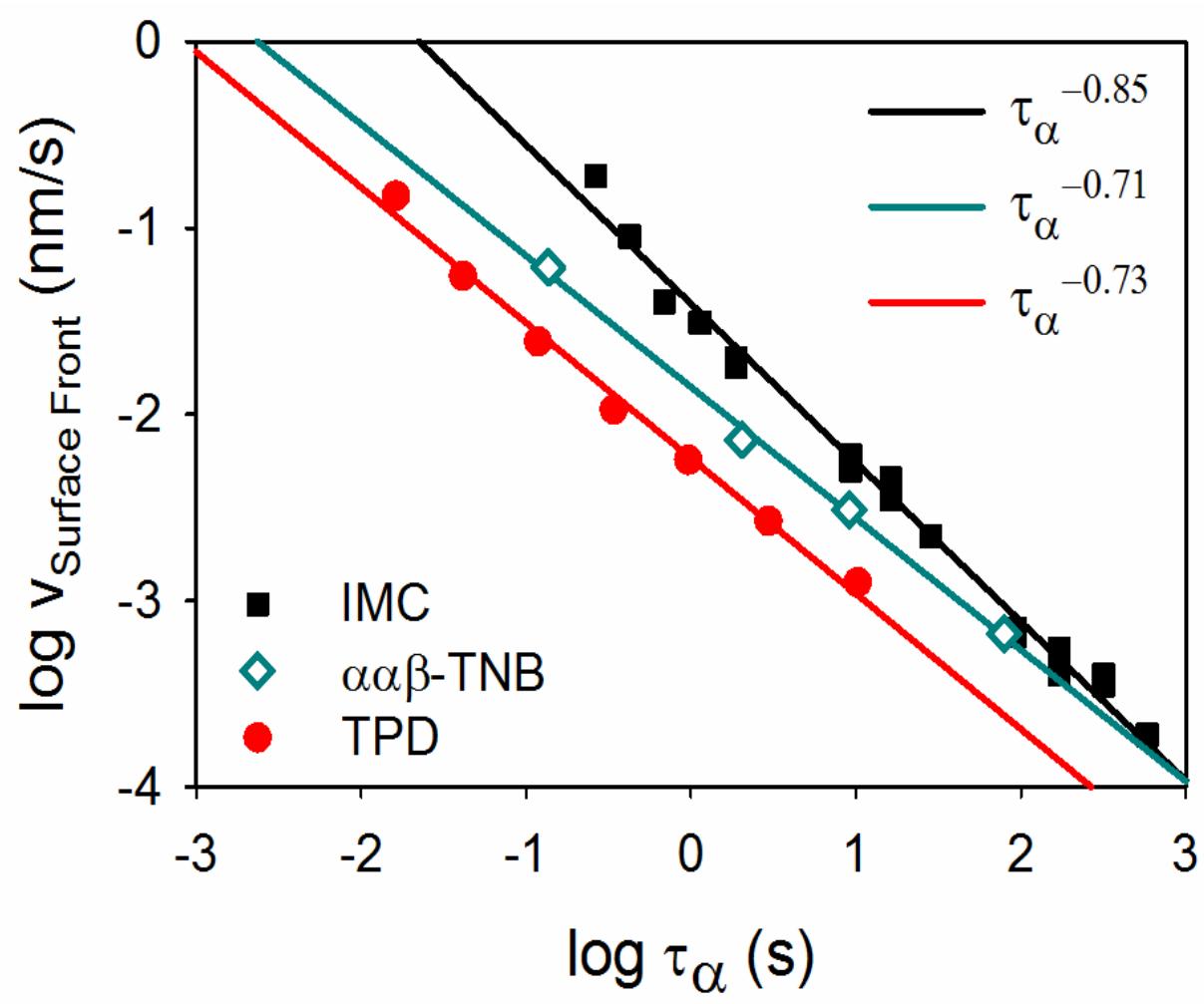

FIG. 7. Surface-initiated transformation front velocities for stable glasses of three molecules as a function of the liquid structural relaxation time at the annealing temperature. In each case, samples were vapor-deposited at 0.85 $T_g$. Indomethacin (IMC) and ααβ-trisnaphthylbenzene (ααβ-TNB) data are previously reported.[32]

Using a high-throughput annealing protocol, we find that the temperature dependence of the transformation front velocity is independent of substrate temperature and is the same for fronts initiated at the free surface and the substrate (Figure 5). A possible explanation for why all the transformation fronts for many different stable glasses have the same activation energy is that all the different glasses transform into the same supercooled liquid with the same mobility. This provides further evidence that the mobility of the supercooled liquid influences the front velocity.

*Influence of Substrate Temperature.* In Figure 8a, the transformation front velocities for TPD glasses are scaled to the mobility of the supercooled liquid at the annealing temperature and compared. In constructing the ordinate, we divide the front velocities by $\tau_\alpha^{-0.73}$, as suggested by Figure 7; this is approximately equivalent to an activation energy of 380 kJ/mol, in agreement with Figure 5. When this scaling is applied, the front velocities at many different annealing temperatures collapse onto a single curve to a good approximation. The successful data collapse shows that the influences of substrate temperature and annealing temperature on the propagation front velocity are independent. The curve shown in Figure 8a expresses the complete dependence of the front velocity on substrate temperature. In this section we consider why glasses deposited at different substrate temperatures transform at different rates.

Figure 8a shows that the substrate temperature during deposition has a large impact on the transformation front velocity for TPD glasses, independent of annealing temperature. The front velocity can vary by more than an order of magnitude with substrate temperature, similar to published results for the model glassformer indomethacin (Figure 4). This significant dependence of front velocity upon substrate temperature has not yet been captured by calculations using kinetic Ising models or RFOT theory.[34,36] A published calculation of front velocities for stable glasses of ααβ-trisnaphthylbenzene showed no systematic dependence on the fictive temperature, which specifies the stability of the glass, over a 10 K range.[36] Based on aging experiments on ordinary glasses,[49] it would be reasonable if denser glasses exhibited lower molecular mobility and lower transformation front velocities. The densities of vapor-deposited TPD glasses are shown in Figure 8b. A comparison between the two panels of Figure 8 illustrates that the front velocities for TPD are partially correlated with density. TPD glasses with the greatest densities ($T_{Substrate}$ in the range 260 to 285 K ) have the lowest front velocities. However, glasses of equivalent densities can show quite different front velocities. Glasses with $T_{Substrate}$ greater than 285 K transform more slowly than glasses of equivalent density when $T_{Substrate}$ is less than 285 K. A similar imperfect correlation between propagation front velocity and glass density has been reported for indomethacin.[27]

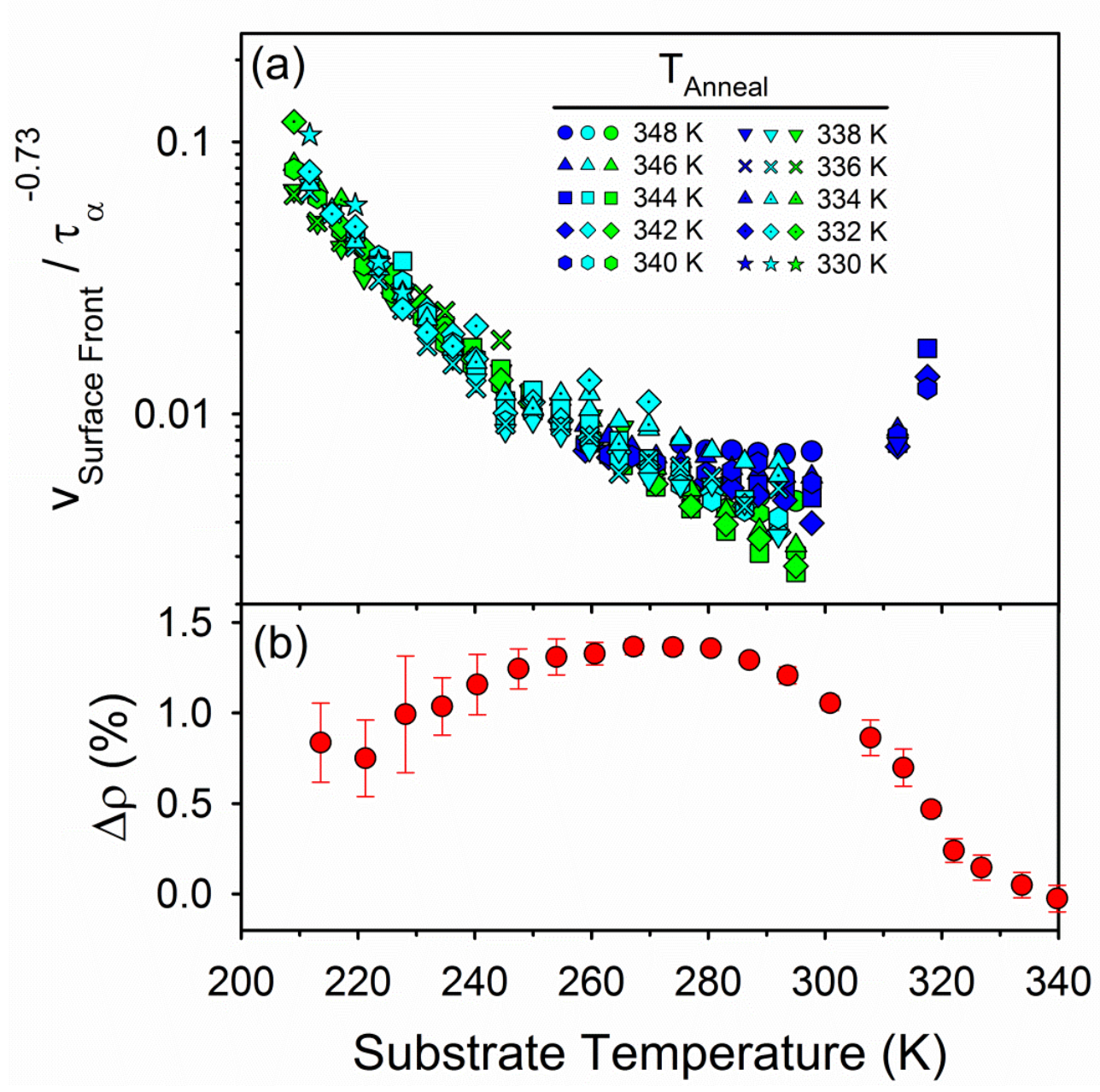

FIG. 8. (a) Surface-initiated transformation front velocities for annealing temperatures ranging from 332 to 348 K scaled by $\tau_\alpha^{-0.73}$ at the annealing temperature, for glasses prepared at a wide range of substrate temperatures. Scaled front velocities are approximately the same for all annealing temperatures and show the same substrate temperatures dependence. Different colors represent different samples, and individual symbols are used for each annealing temperature. (b) Density of the vapor-deposited glasses of TPD relative to the ordinary glass prepared by cooling the liquid at 1K/min, for glasses prepared at different substrate temperatures. Data are from this work and reference 17. Each data point is the average across multiple samples for 2 to 10 glasses with similar substrate temperatures, with the density difference measured at room temperature. Error bars give the 90% confidence interval.

## *C. Stability in Organic Electronics*

Thermal stability in vapor-deposited glasses is important for maintaining high performance in organic electronics. On annealing, vapor-deposited active layers can form pinholes, crystallize, lose mechanical integrity, and lose anisotropy. These changes in the properties of the organic layers can cause organic electronic devices to fail or reduce their efficiency. Understanding how the properties of active layers evolve during annealing could aid in designing organic electronics with improved thermal stability and extended lifetimes.

The thermal stability of active layers in many organic electronics devices may be controlled by transformation front propagation. Many organic molecules have been shown to form stable glasses, including molecules used in organic devices[17], and stable glass formation occurs over a wide range of substrate temperatures; thus many active layers may be stable glasses. The front transformation mechanism would potentially be applicable to any active layer that forms a stable glass, and the front velocity allows for a rough estimate of the thermal stability for such an active layer. Front velocity can be estimated based on the annealing temperature and the substrate temperature of the glass, as shown in Figures 7 and 8. While this method may not be quantitatively accurate, particularly since it is unknown how many fronts will be initiated in an active layer, it still provides guidance for how deposition conditions might change the thermal stability of an active layer.

Adachi and coworkers have also used ellipsometry to investigate the transformation mechanism of glasses composed of molecules used in organic electronics.[23] Their data were better described with a graded mobility model rather than with a transformation front model like the one used here. Their graded model described the vapor-deposited films as three layers with variable thicknesses where each layer is some mixture of the pre- and post-transformation material. While they did not observe a sharp transformation front, they did find that the vapor-deposited glasses transformed first at the surface and later in the film interior. This feature appears robust in the transformation of thin films of vapor-deposited stable glasses, although the generality of the graded transformation mechanism requires further investigation.

Understanding the mechanism of thin film transformation could aid in designing active layers in organic electronics with extended lifetimes and improved thermal stability. Replacing free surfaces with a “capping” layer, a vapor-deposited glassy film of a high $T_g$, low mobility material, can suppress transformation fronts at interfaces as recently demonstrated.[33] The capping layer eliminates the high mobility material that would otherwise initiate a front. Transformation may still eventually initiate from such an interface but it is possible that interface modification, perhaps by tailoring deposition to control the interface breadth between two adjacent layers, might provide further stability. Understanding the transformation mechanism offers ideas for suppressing transformation fronts and designing organic electronics with enhanced lifetimes.

## V. CONCLUDING REMARKS

In this work, we show that vapor-deposited glasses of TPD with high thermal stability transform into the supercooled liquid via fronts propagating with constant velocity. This is the first illustration that a compound used as an active layer in organic electronics transforms via propagating fronts. Transformation fronts in TPD stable glasses have similar velocities and a similar dependence on $T_{Substrate}$ as previously studied model glass formers. Front velocity can vary by more than an order of magnitude for TPD glasses prepared with different substrate temperatures. By using a new high-throughput annealing protocol, the activation energies of the transformation fronts are measured and found to be independent of substrate temperature. We find that the effect of annealing temperature and the substrate temperature on the front velocity are independent. Density is imperfectly correlated with front velocity but may be a useful proxy for front velocity since the densest glasses are the most stable and have the lowest front velocities.

An improved understanding of the thermal stability of an organic semiconductor could lead to better design of organic electronics. We expect that many of the amorphous active layers utilized in organic devices are stable glasses that transform via propagating fronts. For such systems, the results presented here provide a means of predicting which deposition conditions will result in devices with the highest thermal stability. The role of device interfaces in initiating front propagation deserves more study. It is possible that eliminating high mobility interfaces bordering an active layer may improve device lifetime.

## ACKNOWLEDGEMENTS

We thank Shakeel Dalal for helpful discussions. The ellipsometry work (D.M.W., M.D.E.) was supported by the U.S. Department of Energy, Office of Basic Energy Sciences, Division of Materials Sciences and Engineering, award DE-SC0002161. The dielectric relaxation experiments (R.R.) were supported by NSF CHE-1265737.

## REFERENCES

[1] C.A. Angell, Science **267**, 1924 (1995).

[2] C.A. Angell, K.L. Ngai, G.B. McKenna, P.F. McMillan, and S.W. Martin, J. Appl. Phys. **88**, 3113 (2000).

[3] D. Yokoyama, J. Mater. Chem. **21**, 19187 (2011).

[4] K. Ishii and H. Nakayama, Phys. Chem. Chem. Phys. **16**, 12073 (2014).

[5] I. Lyubimov, M.D. Ediger, and J.J. de Pablo, J. Chem. Phys. **139**, 144505 (2013).

[6] S.F. Swallen, K.L. Kearns, M.K. Mapes, Y.S. Kim, R.J. McMahon, M.D. Ediger, T. Wu, L. Yu, and S. Satija, Science **315**, 353 (2007).

[7] K. Dawson, L. Zhu, L.A. Kopff, R.J. McMahon, L. Yu, and M.D. Ediger, J. Phys. Chem. Lett. **2**, 2683 (2011).

[8] K. Ishii, H. Nakayama, R. Moriyama, and Y. Yokoyama, Bull. Chem. Soc. Jpn. **82**, 1240 (2009).

[9] H. Nakayama, K. Omori, K. Ino-u-e, and K. Ishii, J. Phys. Chem. B **117**, 10311 (2013).

[10] S.S. Dalal, A. Sepúlveda, G.K. Pribil, Z. Fakhraai, and M.D. Ediger, J. Chem. Phys. **136**, 204501 (2012).

[11] S.S. Dalal and M.D. Ediger, J. Phys. Chem. Lett. **3**, 1229 (2012).

[12] E. León-Gutierrez, G. Garcia, M.T. Clavaguera-Mora, and J. Rodríguez-Viejo, Thermochim. Acta **492**, 51 (2009).

[13] S.L.L.M. Ramos, M. Oguni, K. Ishii, and H. Nakayama, J. Phys. Chem. B **115**, 14327 (2011).

[14] C. Rodríguez-Tinoco, M. Gonzalez-Silveira, J. Ràfols-Ribé, G. Garcia, and J. Rodríguez-Viejo, J. Non. Cryst. Solids **407**, 256 (2015).

[15] L. Zhu and L. Yu, Chem. Phys. Lett. **499**, 62 (2010).

[16] P.-H. Lin, I. Lyubimov, L. Yu, M.D. Ediger, and J.J. de Pablo, J. Chem. Phys. **140**, 204504 (2014).

[17] S.S. Dalal, D.M. Walters, I. Lyubimov, J.J. de Pablo, and M.D. Ediger, (in press) (2015).

[18] K.J. Dawson, L. Zhu, L. Yu, and M.D. Ediger, J. Phys. Chem. B **115**, 455 (2011).

[19] K. Dawson, L.A. Kopff, L. Zhu, R.J. McMahon, L. Yu, R. Richert, and M.D. Ediger, J. Chem. Phys. **136**, 094505 (2012).

[20] H. Aziz and Z.D. Popovic, Chem. Mater. **16**, 4522 (2004).

[21] T.Y.-H. Lee, Q. Wang, J.U. Wallace, and S.H. Chen, J. Mater. Chem. **22**, 23175 (2012).

[22] G. Nenna, M. Barra, A. Cassinese, R. Miscioscia, T. Fasolino, P. Tassini, C. Minarini, and D. della Sala, J. Appl. Phys. **105**, 123511 (2009).

[23] T. Komino, H. Nomura, M. Yahiro, and C. Adachi, J. Phys. Chem. C **116**, 11584 (2012).

[24] S.F. Swallen, K. Traynor, R.J. McMahon, and M.D. Ediger, Phys. Rev. Lett. **102**, 065503 (2009).

[25] K.L. Kearns, M.D. Ediger, H. Huth, and C. Schick, J. Phys. Chem. Lett. **1**, 388 (2010).

[26] A. Sepúlveda, M. Tylinski, A. Guiseppi-Elie, R. Richert, and M.D. Ediger, Phys. Rev. Lett. **113**, 045901 (2014).

[27] S.S. Dalal and M.D. Ediger, J. Phys. Chem. B **119**, 3875 (2015).

[28] M. Ahrenberg, Y.Z. Chua, K.R. Whitaker, H. Huth, M.D. Ediger, and C. Schick, J. Chem. Phys. **138**, 024501 (2013).

[29] D. Bhattacharya and V. Sadtchenko, J. Chem. Phys. **141**, 094502 (2014).

[30] C. Rodríguez-Tinoco, M. Gonzalez-Silveira, J. Ràfols-Ribé, A.F. Lopeandía, M.T. Clavaguera-Mora, and J. Rodríguez-Viejo, J. Phys. Chem. B **118**, 10795 (2014).

[31] Z. Chen, A. Sepúlveda, M.D. Ediger, and R. Richert, J. Chem. Phys. **138**, 12A519 (2013).

[32] A. Sepúlveda, S.F. Swallen, L.A. Kopff, R.J. McMahon, and M.D. Ediger, J. Chem. Phys. **137**, 204508 (2012).

[33] A. Sepúlveda, S.F. Swallen, and M.D. Ediger, J. Chem. Phys. **138**, 12A517 (2013).

[34] S. Léonard and P. Harrowell, J. Chem. Phys. **133**, 244502 (2010).

[35] P.G. Wolynes, Proc. Natl. Acad. Sci. **106**, 1353 (2009).

[36] A. Wisitsorasak and P.G. Wolynes, Phys. Rev. E **88**, 022308 (2013).

[37] C.W. Brian and L. Yu, J. Phys. Chem. A **117**, 13303 (2013).

[38] C. Adachi, T. Tsutsui, and S. Saito, Appl. Phys. Lett. **55**, 1489 (1989).

[39] S.S. Dalal, Z. Fakhraai, and M.D. Ediger, J. Phys. Chem. B **117**, 15415 (2013).

[40] (SPACE FOR SUPPLEMENTAL MATERIAL), (2015).

[41] K.L. Kearns, S.F. Swallen, M.D. Ediger, T. Wu, and L. Yu, J. Chem. Phys. **127**, 154702 (2007).

[42] R. Richert, K. Duvvuri, and L.-T. Duong, J. Chem. Phys. **118**, 1828 (2003).

[43] H. Vogel, Phys. Z. **22**, 645 (1921).

[44] G.S. Fulcher, J. Am. Ceram. Soc. **8**, 339 (1925).

[45] G. Tammann and W. Hesse, Z. Anorg. Allg. Chem. **156**, 245 (1924).

[46] G. Fredrickson and H. Andersen, Phys. Rev. Lett. **53**, 1244 (1984).

[47] D. Chandler and J.P. Garrahan, Annu. Rev. Phys. Chem. **61**, 191 (2010).

[48] A.S. Keys, L.O. Hedges, J.P. Garrahan, S.C. Glotzer, and D. Chandler, Phys. Rev. X **1**, 021013 (2011).

[49] S.L. Simon and P. Bernazzani, J. Non. Cryst. Solids **352**, 4763 (2006).

Supplemental Material for:

# Thermal Stability of Vapor-Deposited Stable Glasses of an Organic Semiconductor

Diane M. Walters[†], Ranko Richert[‡], M. D. Ediger[†*a]

[†]*Department of Chemistry, University of Wisconsin – Madison, Madison, Wisconsin 53706, USA*

[‡]*Department of Chemistry and Biochemistry, Arizona State University, Tempe, Arizona 85287, USA*

## Optical Constants of Vapor-Deposited TPD Glasses:

The optical constants of the vapor-deposited TPD glasses were measured using spectroscopic ellipsometry as specified in the experimental section of the text. The as-deposited glasses are birefringent and dichroic, and so have different refractive indexes and the extinction coefficients in and out of the plane of the substrate. The optical properties for as-deposited TPD glasses prepared with a wide range substrate temperatures are reported in Figure S1 below. Additionally, Figure S1 demonstrates that the optical constants of independently prepared samples, represented by different symbols, are highly reproducible.

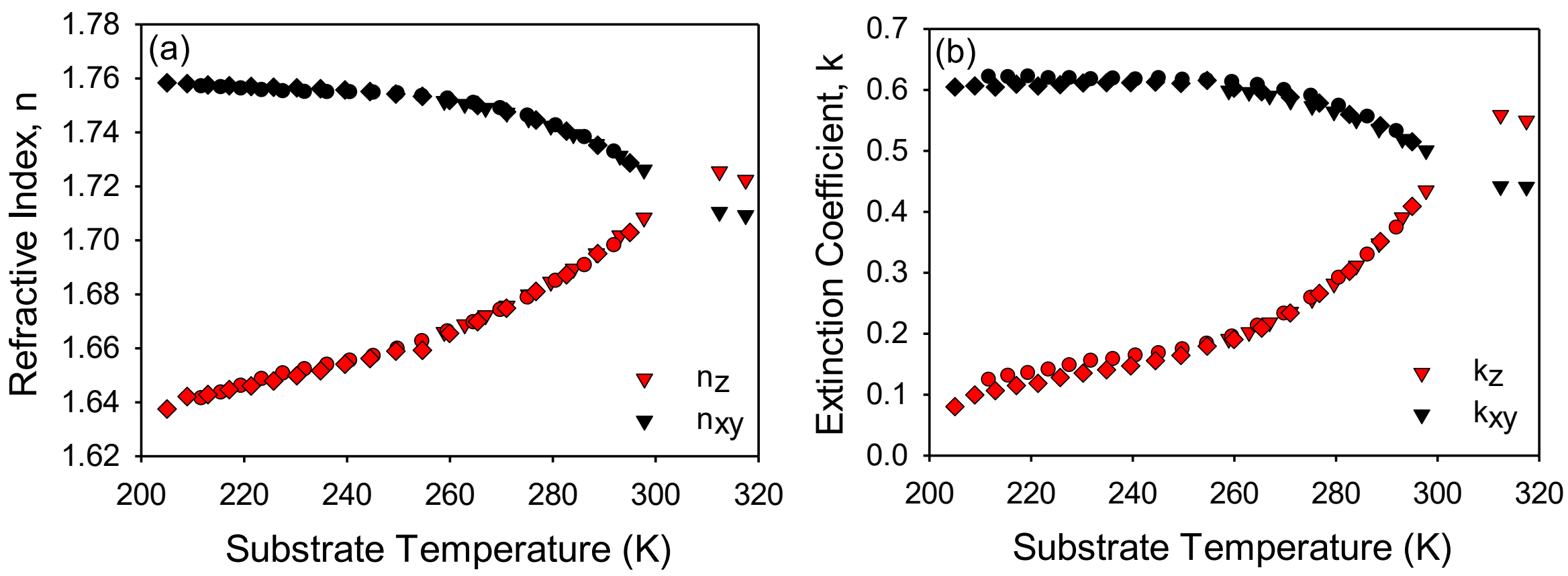

FIG. S1. The optical properties of as-deposited TPD glasses prepared at different substrate temperatures. Different symbols represent independently prepared samples. (a) Refractive indexes out of the plane ($n_z$) and in the plane ($n_{xy}$) of the substrate. (b) Extinction coefficients out of the plane ($k_z$) and in the plane ($k_{xy}$) of the substrate.

[a] Corresponding Author: Contact at ediger@chem.wisc.edu. Phone: 608-262-7273